\documentstyle[aps]{revtex}
\begin{document}
\input{epsf}
\title{Measurements of Neutrons in 11.5 A GeV/c Au~+~Pb Heavy-Ion Collisions}
\author{
T.A. Armstrong                \unskip,$^{(8,\ast)}$
K.N. Barish                   \unskip,$^{(3)}$
S. Batsouli                   \unskip,$^{(13)}$
S.J. Bennett                  \unskip,$^{(12)}$
A. Chikanian                  \unskip,$^{(13)}$
S.D. Coe                      \unskip,$^{(13,\dag)}$
T.M. Cormier                  \unskip,$^{(12)}$
R. Davies                     \unskip,$^{(9,\ddag)}$
C.B. Dover                    \unskip,$^{(1,\S)}$
P. Fachini                    \unskip,$^{(12)}$
B. Fadem                      \unskip,$^{(5)}$
L.E. Finch                    \unskip,$^{(13)}$
N.K. George                   \unskip,$^{(13)}$
S.V. Greene                   \unskip,$^{(11)}$
P. Haridas                    \unskip,$^{(7)}$
J.C. Hill                     \unskip,$^{(5)}$
A.S. Hirsch                   \unskip,$^{(9)}$
R. Hoversten                  \unskip,$^{(5)}$
H.Z. Huang                    \unskip,$^{(2)}$
B.S. Kumar                    \unskip,$^{(13,\|)}$
T. Lainis	              \unskip,$^{(10)}$
J.G. Lajoie                   \unskip,$^{(5)}$
Q. Li                         \unskip,$^{(12)}$
B. Libby                      \unskip,$^{(5,\P\P)}$
R.D. Majka                    \unskip,$^{(13)}$
T.E. Miller                   \unskip,$^{(11)}$
M.G. Munhoz                   \unskip,$^{(12)}$
J.L. Nagle                    \unskip,$^{(4)}$
I.A. Pless                    \unskip,$^{(7)}$
J.K. Pope                     \unskip,$^{(13,\ast\ast)}$
N.T. Porile                   \unskip,$^{(9)}$
C.A. Pruneau                  \unskip,$^{(12)}$
M.S.Z. Rabin                  \unskip,$^{(6)}$
J.D. Reid                     \unskip,$^{(11)}$\\
A. Rimai                      \unskip,$^{(9,\dag\dag)}$
A. Rose                       \unskip,$^{(11)}$
F.S. Rotondo                  \unskip,$^{(13,\ddag\ddag)}$
J. Sandweiss                  \unskip,$^{(13)}$
R.P. Scharenberg              \unskip,$^{(9)}$
A.J. Slaughter                \unskip,$^{(13)}$
G.A. Smith                    \unskip,$^{(8)}$
M.L. Tincknell                \unskip,$^{(9,\S\S)}$
W.S. Toothacker               \unskip,$^{(8)}$
G. Van Buren                  \unskip,$^{(2)}$
F.K. Wohn                     \unskip,$^{(5)}$
Z. Xu                         \unskip,$^{(13)}$}
\address{\centerline{(The E864 Collaboration)}}
\address{  $^{(1)}$ Brookhaven National Laboratory, Upton, 
New York 11973 \break
  $^{(2)}$ University of California at Los Angeles, Los Angeles, 
California 90095 \break  
  $^{(3)}$ University of California at Riverside, Riverside, 
California 92521 \break
  $^{(4)}$ Columbia University, New York 10027 \break
  $^{(5)}$ Iowa State University, Ames, Iowa 50011 \break 
  $^{(6)}$ University of Massachusetts, Amherst, Massachusetts 01003 \break 
  $^{(7)}$ Massachusetts Institute of Technology, Cambridge, 
Massachusetts 02139 \break 
  $^{(8)}$ Pennsylvania State University, University Park, 
Pennsylvania 16802 \break 
  $^{(9)}$ Purdue University, West Lafayette, Indiana 47907 \break 
  $^{(10)}$ United States Military Academy, West Point \break
  $^{(11)}$ Vanderbilt University, Nashville, Tennessee 37235 \break 
  $^{(12)}$ Wayne State University, Detroit, Michigan 48201 \break 
  $^{(13)}$ Yale University, New Haven, Connecticut 06520 \break
}

\date{\today}
\maketitle
\begin{abstract}
We present measurements from Brookhaven AGS Experiment 864 of neutron
invariant multiplicity in 11.5 A GeV/c Au~+~Pb collisions.  
The measurements span a rapidity range from center
of mass to beam rapidity ($y_{beam}$=3.2) and are presented as a
function of event centrality.  The results are compared with
E864 measurements of proton invariant multiplicity and an average
$n/p$ ratio at hadronic freeze-out of 1.19$\pm$.08 is determined for
the rapidity range $y$=1.6 to $y$=2.4.  We discuss briefly the
implications of this ratio within a simple equilibrium model of
the collision system.
\end{abstract}
\pacs{25.75.-q}

\section{Introduction}
As constituents of the colliding nuclei in a relativistic
Au~+~Pb heavy ion interaction, neutrons carry over
60\% of the incident energy.  Knowledge of the final distribution
of neutrons resulting from these collisions is then important
for the determination of the amount of energy deposited
in the central rapidity region in such collisions.  Because 
the time scale of these collisions is so small (the total duration
until hadronic freeze out is believed to be on the order  
10 fm/c \cite{RQMD}), the collision dynamics are dictated by the strong 
interaction. It is then reasonable to assume that the behaviour
of neutrons in the collision should closely parallel that of protons which
have been extensively measured for similar collision systems and 
energies \cite{E866,E877,Nigel}.
Indeed, in the absence of neutron data this has been widely 
assumed in the calculation of
light nuclei coalescence parameters (sometimes
with the explicit assumption that the neutron to proton ratio 
available for coalescence is the primordial ratio) \cite{jamie_col,sato}. 

It is also widely assumed that the neutron to proton ratio at hadronic 
freeze-out should show significant equilibration from the initial 
Au~+~Pb ratio of 1.52:1.  Measurements of nucleons in Au~+~Au collisions
at lower energy \cite{sis} do in fact exhibit this.
This equilibration is expected to be enhanced at AGS energies as a result of 
a large amount of strong
resonance production in the collision region which should have the
effect of speeding the system toward chemical equilibrium and so transferring 
some of the initial isospin imbalance to an excess of 
$\pi^{-}$ over $\pi^{+}$.  

\section{Experiment 864}

\subsection{The E864 Spectrometer}

\begin{figure}
\centering\leavevmode\epsfbox{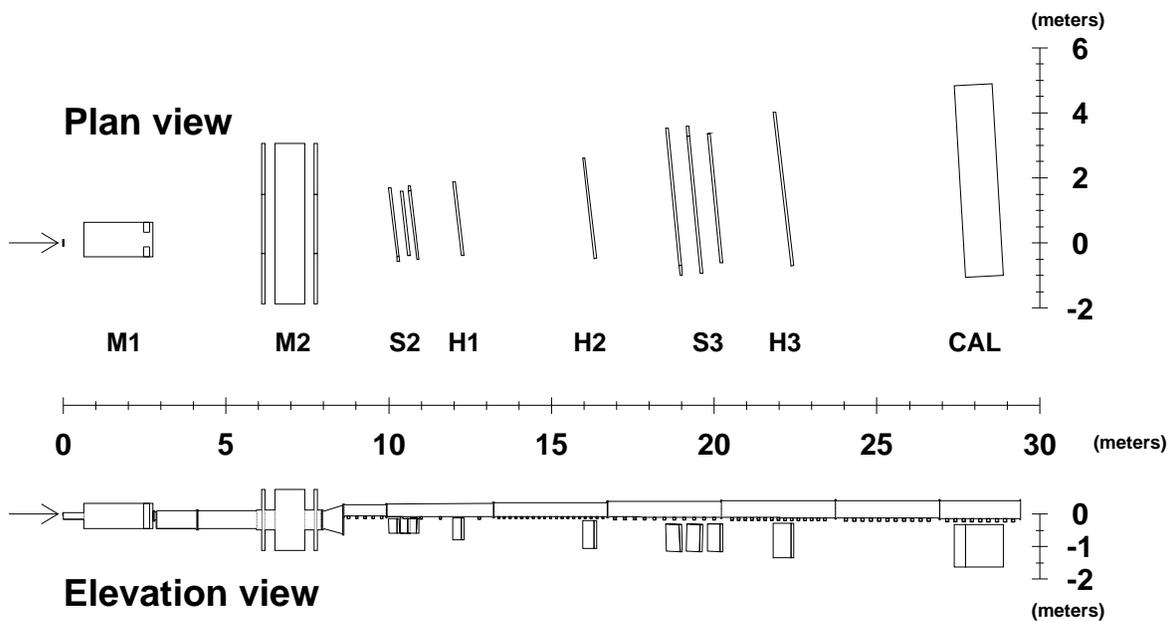}
\caption{The E864 spectrometer in plan and elevation views, showing the
dipole magnets (M1 and M2), hodoscopes (H1, H2, and H3), straw tube arrays 
(S1 and S2) and hadronic calorimeter (CAL).  The vacuum chamber is not 
shown in the plan view. }
\label{fig:john_ap}
\end{figure}

BNL Experiment 864 is an open geometry, high data rate spectrometer that 
was designed chiefly to search for cold strange quark matter 
(strangelets) which may be produced in heavy ion collisions.  Plan
and elevation views of the spectrometer are shown in Figure~\ref{fig:john_ap}.
A thorough description of the apparatus is provided in \cite{bignim}.

A beam of gold ions with momentum 11.5 A GeV/c is incident on a fixed lead
target.  The interaction products then travel downstream through two
dipole magnets M1 and M2.  A collimator inside of M1 defines the experimental
acceptance; for neutral particles this is -32 mr to 114 mr in the 
horizontal and -17 mr to -51.3 mr in the vertical.  

The charged particle tracking system consists of three hodoscope scintillator
walls (H1, H2, and H3) and two straw tube stations (S2 and S3).  The hodoscopes
provide for each charged particle hit a measurement 
of time, charge, and position.  
This information is then used to build track candidates which are either 
rejected or confirmed and further refined by straw tube position 
information.  With knowledge of the fields in M1 and M2, tracked particles 
are then identified by mass computed through 
rigidity, charge, and velocity with the assumption that the tracks originate
from the target.

\begin{figure}
\centering\epsfxsize=5in \epsfysize=6in \leavevmode\epsfbox{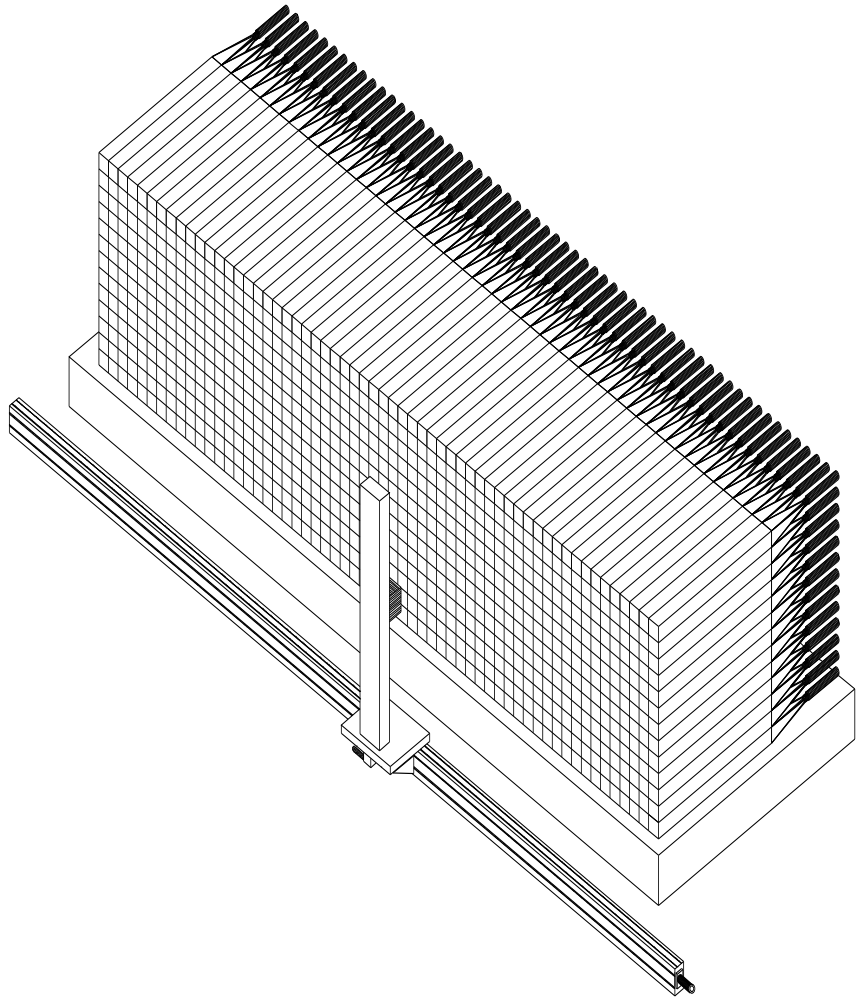}
\caption{The E864 hadronic sampling calorimeter; an array of 58x13 towers.  
The active material is scintillating fiber running in small strips lengthwise 
down each lead tower in a spaghetti design.  Also pictured is the Cobalt 60 
calibration system which is used for gain matching among the towers.}
\label{fig:whole_calo}
\end{figure}

At the downstream end of the apparatus is the E864 hadronic calorimeter 
which is 
crucial to the neutral particle analyses.  The calorimeter (see Figure 
~\ref{fig:whole_calo}) consists of an array of 58x13 towers, each 10cm x 10cm 
on the front face and 117cm long.  This lead/scintillator sampling 
calorimeter is of a spaghetti design with scintillating fibers running 
lengthwise down each calorimeter tower giving a total lead to scintillator 
ratio of 4.55:1 by volume.  The calorimeter has excellent resolution for 
hadronic showers in energy ($\sigma_{E}/E = .34/(\sqrt{E})+.035$ for $E$ in
units of GeV) and time ($\sigma_{t} \approx$ 400 ps) and is described 
in detail in \cite{claudecal}.

Collision centrality is defined in E864 through a measurement of charged 
particle multiplicity.  The E864 multiplicity counter \cite{beam} is an annular
piece of scintillator placed around the beam pipe 13 cm downstream of the 
target that subtends an angular range from 16.6$^{o}$ to 45.0$^{o}$.  The 
annulus is separated into four quadrants, each of which is viewed by a 
photomultiplier tube.  The sum of the integrated charge signal from the four 
quadrants is proportional to the charged particle multiplicity of the collision
and is used to define event centrality. 

\subsection{Neutron analysis}

We measure the invariant multiplicity of neutrons by dividing momentum
space into bins in rapidity and transverse momentum of size $\Delta y$ 
by $\Delta p_{T}$.  In terms of the actual experimental quantities, we then
have the invariant multiplicity in a momentum bin with average transverse 
momentum $<p_{T}>$ as
\begin{equation}
\frac{1}{2\pi p_{T}} \frac{d^{2}N}{dydp_{T}} = \frac{1}{2\pi <p_{T}> \Delta
y \Delta p_{T}} \frac{N_{counts}}{N_{events}} \frac{1}
{\epsilon_{ACC}(y,p_{T}) \times
\epsilon_{REC}(y,p_{T}) }
\end{equation} 

Here $N_{counts}$ is the number of neutrons reconstructed in our calorimeter 
analysis of $N_{events}$.  $\epsilon_{ACC}(y,p_{T})$ is the geometric 
acceptance for neutrons in our apparatus and $\epsilon_{REC}(y,p_{T})$ is
the efficiency for reconstructing with our analysis algorithm those neutrons 
which are accepted.    

The first step in determining $N_{counts}$ is to identify all those 
calorimeter towers in a given event
which are peak towers.  We define a peak tower as a tower which has more 
energy deposited than any of its 8 neighbors.  For each peak tower, 
we define the corresponding 
energy shower as including all towers in a 3x3 grid centered on the peak 
tower.  A 3x3 array is used because the improvement in energy resolution 
obtained by using a 5x5 grid is 
slight and the resulting contamination is much larger.  Approximately 
90\% of the shower energy for a neutron with a kinetic energy of 6 GeV is
contained in a 3x3 grid.

Each of these energy showers is then put through the following series
of contamination cuts:
\begin{itemize}
\item There must be no charged particle 
track found using only the hodoscopes
which points to any of the nine towers in the shower.  With the 
dipole fields in M1 and M2 set to 1.5 Tesla (the fields are aligned
in the same direction), most charged particles are swept
out of the neutron fiducial region so that the ratio of proton 
hits to neutrons hits
is approximately 1 to 3 with some variation as a function of position.
Charged pions, kaons, and deuterons are light and/or rare enough that
calorimeter hits due to these species are suppressed by at least another 
order of magnitude.  

Of the charged particle 
peaks in the neutral fiducial region
of the calorimeter, we estimate from monte carlo simulations that 
81\% are rejected by this method compared with 6\% of neutron peaks.
\item There must be no energy peak larger than some minimum energy, $E_{PK}$, 
in the square of 16 towers which borders the shower.  Values of $E_{PK}$ used
in the analysis vary from 1.5 to 2.5 GeV as a function of rapidity.
\item A cut is made on the ratio, $R_{5x5/3x3}$, of total 
energy in 25 towers around the peak to total energy in 9 towers around the peak.
The maximum allowable value of $R_{5x5/3x3}$ in the analysis ranged from
1.7 to 2.5.  The cut values for $E_{PK}$ and $R_{5x5/3x3}$ were chosen by 
observation of the value
of these quantities for those showers which were designated as clean by
other contamination cuts, with consideration given both to the level of 
background present as a function of rapidity and to keeping the
efficiency as high as reasonably possible.   
\item Each tower of the shower which has a nonzero time must show 
agreement within a time window, $t_{max}$, 
with the peak tower.  Values of $t_{max}$ used were 1.6 ns for bins of
$y$ = 1.7 and 1.75 ns elsewhere.  Side tower time resolutions in 
neutron showers were approximately 500 ps (not gaussian) with some 
variation as a function of energy.  This cut was adjusted to be
95\% efficient for isolated neutron showers at rapidity 1.9 and above.   
\item In bins of rapidity 2.5 and greater, a clear separation can 
be seen between neutrons and photons on a plot of shower mass versus 
percentage of shower energy in the peak tower, indicating that neutron 
showers are much wider than photon showers.  For these rapidities, 
we place a cut on the 
ratio of energy in the peak tower to energy in the 9 tower sum to 
reduce contamination from
photons; rejecting showers for which this ratio is
larger than 0.83.
\item Finally, the shower energy profile is compared with the energy profiles 
of several hundred thousand isolated proton showers which span the full range
of incident angles and front face hit positions and most of the rapidity 
range of neutrons incident on the calorimeter.  

The fraction of shower energy in each of the nine towers is calculated
and rounded to the nearest 5\%.  This set of nine fractions is then
compared with the set of fractions for each of these isolated proton showers. 
If fewer than two matching sets of fractions are found, the shower
is discarded.

\end{itemize}

For those showers which survive the cuts listed above, 
mass is calculated from the peak tower time,
nine tower energy sum, and shower position as 
$m=E_{sum}/(\gamma_{peak}-1)$.  A momentum is also assigned
to the shower assuming a neutron mass and using energy and time measurements 
weighted according to their errors.

\begin{figure}
\centering\epsfxsize=5in \epsfysize=6in \leavevmode \epsfbox{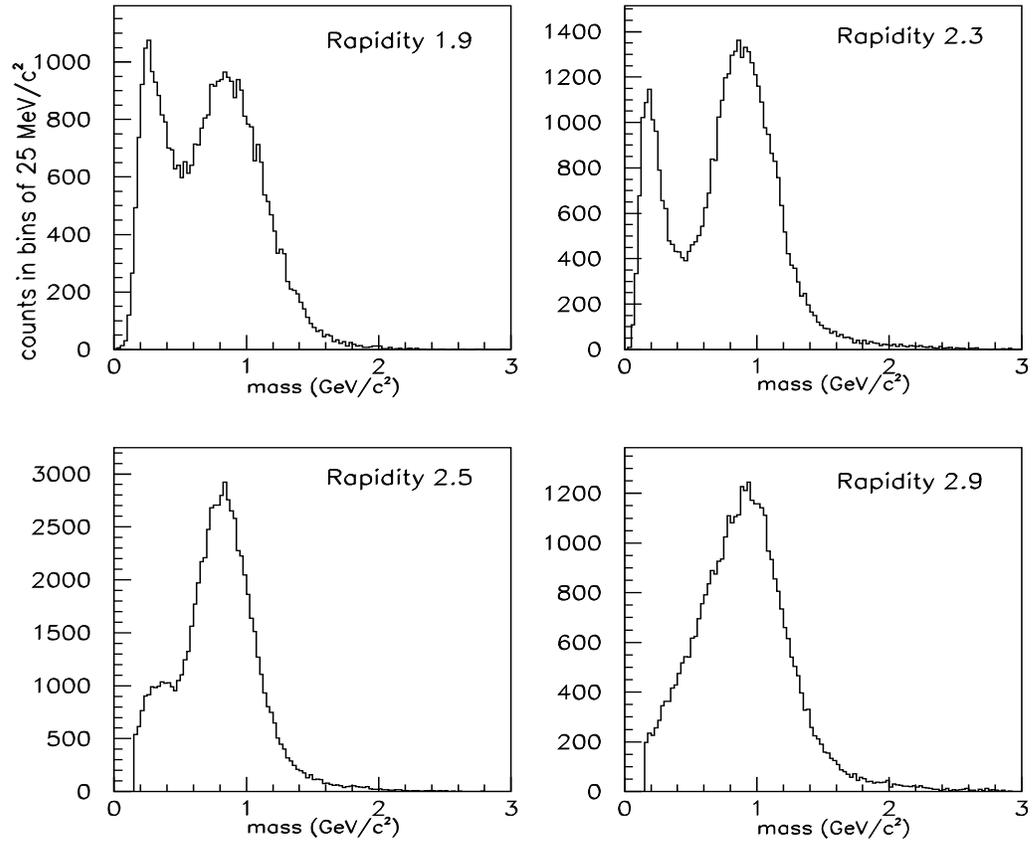}
\caption{Sample mass histograms in four different momentum bins:  $y$ = 1.8-2.0 
(200$<p_{T}<$250 MeV/c), $y$ = 2.2-2.4 (350$<p_{T}<$400 MeV/c),
$y$ = 2.4-2.6 (250$<p_{T}<$300 MeV/c) and  $y$ = 2.8-3.0 
(400$<p_{T}<$450 MeV/c). Beam rapidity is 3.2, and neutron 
measurements in this paper are reported from rapidity 
1.6 up to 3.2.  These four histograms represent the nature 
of the neutron mass 
signal and low mass background as a function of rapidity.}
\label{fig:four_bins_2}
\end{figure}

Momentum space is then divided into bins of 50 MeV/c in 
$p_{T}$ by .2 units in $y$.  For each
bin we make a mass plot as shown in Figure~\ref{fig:four_bins_2}.  There is a 
background at low mass which is clearly evident for rapidities of 2.3 
and below.  This background is predicted qualitatively by our GEANT
detector simulations as a mixture of scattered photons and hadrons, 
but the predicted level
is less by a factor of four or more than the level 
seen in the data.  We believe that
this discrepancy is due to a combination of our modelling of the 
calorimeter time response near threshold being somewhat incorrect
and the absence from our GEANT simulations of certain
downstream geometries which may 
contribute scattered particles to this low energy background.    

\begin{figure}
\centering\epsfxsize=5in \epsfysize=6in \leavevmode \epsfbox{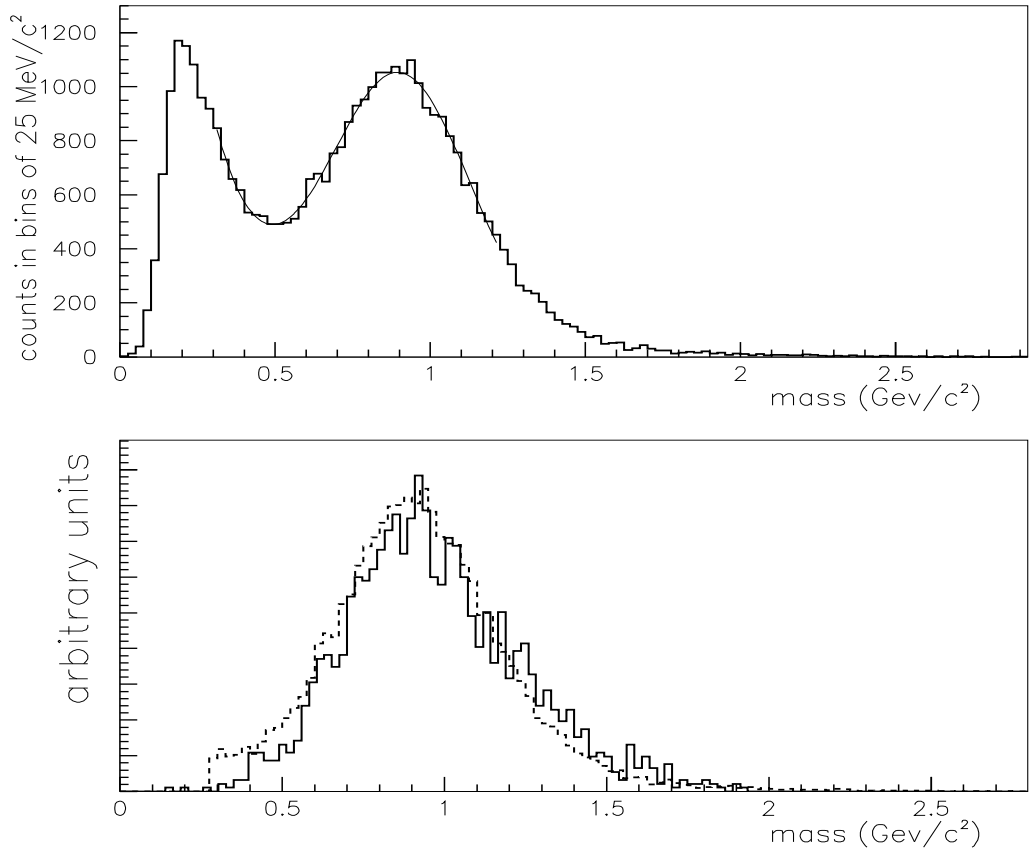}
\caption{An example of background subtraction in one momentum bin; 
1.8 $ < y < $ 2.0, 350 MeV/c $ < p_{T} < $ 400 MeV/c.  The top plot
is simply the mass histogram with a fit to an exponential background
plus a gaussian signal over a limited range of the neutron peak.  
In the 
lower plot, the dashed histogram is this same mass histogram 
with the background subtracted
away according to this parameterization.  The solid histogram which is
overlaid in the lower plot is the result of simulating the calorimeter 
response to neutron 
showers in this momentum bins using identified isolated proton showers.}
\label{fig:zelpap_2}
\end{figure}

We count neutrons in the mass range
from .55 $GeV/c^{2}$ to 1.55 $GeV/c^{2}$ and then subtract out the 
contribution of this background 
according to a parameterization of an exponentially 
decaying background plus a gaussian
signal.  Subtracting away the background in this manner 
leaves a neutron signal shape that agrees 
well with simulations (Figure~\ref{fig:zelpap_2}) in which 
isolated proton showers are overlaid on the calorimeter 
to simulate the calorimeter response to neutron showers 
with the contamination of a heavy-ion
event.  (Note that while the shape agrees well, the energy scale in the 
simulations must often be adjusted by around 5\% to show agreement with the
data; possible systematic error from this effect is dealt with separately as
part of the study of differences in calorimeter response between 
protons and neutrons.)  
This low mass background produces only a small correction 
to the number of neutrons counted; never larger 
than 14\% according to our parameterization.

Two classes of background can produce mass peaks under the neutron peak and are 
subtracted away using monte-carlo simulations.  
The first class is from calorimeter hits by particle species 
other than neutrons.  This includes mainly neutral kaons and protons which are 
missed by the tracking system.  This class amounts to less than 
a 10\% correction to the number of 
counted neutrons in most of the momentum bins in which we measure.  
The second class of background is due to neutrons
which do not come directly from the target but come from inelastic 
scatterings in other parts of the apparatus  
(neutrons which elastically scatter are dealt with as part of
the geometric acceptance calculation).
These are largely from the 
upper edge of the 
collimator which sits approximately
one meter downstream of the target; only scattering sources near the 
target can produce neutrons with time and energy combinations which 
will allow them to fall under the mass peak of neutrons from the target.  
For this background, we have a check on the accuracy of the monte-carlo 
simulations because a similar background is present
for protons.  For protons we can use tracking information to determine
if a track originated in the target or collimator and so we can compare
the monte-carlo predictions of this scattered background to what is 
present in the actual data.  We find agreement to better than 25\%
between the amount of background predicted by detector simulations
using two different input distributions and the background seen in 
the data for the protons.
Corrections for this background are as large as 25\% in central 
collisions near center of mass rapidity and decrease with increasing
rapidity. 

Each of these backgrounds is calculated and subtracted separately in each
$y, p_{T}$ bin:  the backgrounds discussed above are summarized 
in Table~\ref{tab:bkg}.

$\epsilon_{ACC}(y,p_{T})$ is essentially the ratio of the number of 
neutrons which leave the target with momentum inside a given $(y,p_{T})$ 
bin to the number of neutrons which strike the
calorimeter (not including those which are from 
inelastic scattering) with momentum inside that
$(y,p_{T})$ bin.  It is determined simply by a GEANT 
simulation of the experimental apparatus.
The results of the acceptance simulation are largely 
insensitive to the assumed neutron
input distribution, but some sharing between bins does take place 
particularly at large rapidity and transverse momentum.

To determine $\epsilon_{REC}$, we have constructed a library of isolated 
proton showers using the charged tracking system to identify protons and 
contamination cuts both in the calorimeter and from tracking to ensure clean 
showers.  To determine the efficiency for neutron reconstruction in the 
calorimeter as a function of energy and position (or rapidity and transverse 
momentum),  the overall times and energies of these clean proton showers were 
altered while leaving the relative times and energy fractions intact to 
simulate neutron showers.  These fake neutron showers were 
overlaid on complete events in the calorimeter, one per event.  Because in 
this manner we can simulate a shower of a neutron of known
momentum striking the calorimeter in a known position, we can determine the 
efficiency for reconstructing these fake neutron showers and take this
to be our reconstruction efficiency for real neutron showers.  
We find an average efficiency
of approximately 35\% with variations as a function of momentum.  

The efficiencies of the individual analysis cuts are listed 
in Table~\ref{tab:indiecuts}, both for a neutron shower on an 
empty calorimeter and for a neutron shower in a central heavy ion 
event.  These efficiencies vary as a function of momentum and 
so are listed in two different rapidity ranges.

The overall efficiency on an empty calorimeter is on average about 70\% and
higher near central rapidity than near beam rapidity.  The extra factor
of two (from 70\% to 35\%) of loss in efficiency is then due to occupancy
in the calorimeter.  There are on average 10 showers in the neutral 
fiducial region
of the calorimeter with peak energy greater
than 1 GeV in a central event, leading to an overall occupancy of about 15\% of
the towers having an energy of 500 MeV or higher in an average event.  The
occupancy is somewhat greater nearer to the neutral line ($\approx$ 20\%)
and less near the edges ($\approx$ 5\%).  Although the overall occupancy
is smaller for less central events, it is in fact slightly larger near
the neutral line and thus the overall efficiency increases only by a few
percent, with a larger increase at high transverse momentum (away from the
neutral line). 

The efficiencies for finding the neutron showers are crucial numbers in this
analysis, so it is important that the method described above give us
an accurate calculation of these efficiencies.   As a check 
of this method we have repeated the process by 
following essentially the same recipe using protons rather than 
neutrons.  That is, we add a proton shower from our shower library 
(along with fake hits in our other detectors to simulate the corresponding
charged particle track) to a heavy ion event and calculate our efficiency
for finding this fake proton shower (when we find the corresponding fake track).
For protons, we can compare this efficiency to the efficiency 
for finding a real proton shower when we know a real proton hits 
the calorimeter (i.e. we find a proton track in our data). 
We do in fact find that the two methods of calculating the
proton efficiencies agree to within 10\% of one another with
the differences largely explained by inefficiencies in our method
for identifying isolated proton showers (refer to the paragraph following
this one).  With the implicit assumption that proton and neutron shower
energy profiles will be basically indistinguishable at these energies
of a few GeV, we conclude from this study that our method for determining the
neutron efficiencies is sound to within 10\%.  Other
differences in these processes for the protons and neutrons (faking of
a charged particle tracks, slightly different angles of incidence for
protons and neutrons across the calorimeter) have been studied and
are not significant sources of error (\cite{evanthes} 
and \cite{xzbthes,Nigel}, respectively).
 
Two small corrections are then made to the $\epsilon_{REC}$ numbers determined
in this manner.  The first is because we are artificially increasing the 
calorimeter occupancy by adding these fake neutron showers.  We account
for this following reference \cite{stankus} and find that it amounts to 
never more than a 5\% correction for any momentum bin and is significantly 
less over most of the momentum space we measure.  The second is because the 
cuts placed on the proton shower library which were necessary to ensure 
isolated showers result in throwing out a few percent of showers which are not 
significantly contaminated.  In particular, since part of the requirement 
for a clean proton shower is timing agreement among the towers in the shower, 
the set of proton showers does not include the tails of the timing agreement 
distribution.  Thus neutron efficiencies
calculated using proton showers are slightly overestimated.  We determine the 
size of this correction partially by using data from a later run of the 
experiment with an incident beam of protons rather than heavy ions to reduce 
contamination of the proton showers.  We estimate a resulting correction that 
is 2\% at center of mass rapidity and rises to 9\% at beam rapidity.

Sources of possible systematic error which we have quantified include:
\begin{itemize}
\item Error due to the assumed input distributions of neutrons and sharing 
between neighboring bins 
(particularly in transverse momentum) in determination of both 
$\epsilon_{REC}$ and $\epsilon_{ACC}$.  By using alternative input 
distributions, we estimate the size of this effect to be 
approximately 5\% over most of the momentum space which we measure.

\item Possible differences in the calorimeter response to proton and neutron 
showers.  This is quantified by explicitly changing the gains factors
used in the analysis and observing the resulting change in measured yields.  
This adds only 3\% to 5\% systematic error over most of the 
acceptance but becomes larger near edges of our kinematic acceptance.

\item Assumed input distribution for background studies, both for scattered 
neutrons and for particle species other than neutrons.  This adds a 
10\% systematic uncertainty near center of mass rapidity and decreases 
at higher rapidity. 

\item Uncertainty in fit parameters in the subtraction of low mass 
background.  This is estimated to add a maximum systematic error of 
8\% in any given bin and the uncertainty decreases as rapidity increases.

\end{itemize}

The statistical errors are generally dominated by systematics.  We add these 
two types of errors in quadrature and list the total uncertainty in 
each bin along with the measurement in Table~\ref{tab:cnt}.

\section{Results and Discussion}

As in the E864 light nuclei measurements, we divide events
into three centrality classes: 10\% most central, 10-38\% 
central and 38-66\% central.  The centrality is defined by our multiplicity 
detector and is reported in terms of percent of the geometric cross section 
defined by $\sigma = \pi r_{0}^{2}(A_{Au}^{1/3} + A_{Pb}^{1/3})^{2}$ with 
$r_{0}$ = 1.2 fm.

\begin{figure}
\centering\epsfxsize=5in \epsfysize=6in \leavevmode \epsfbox{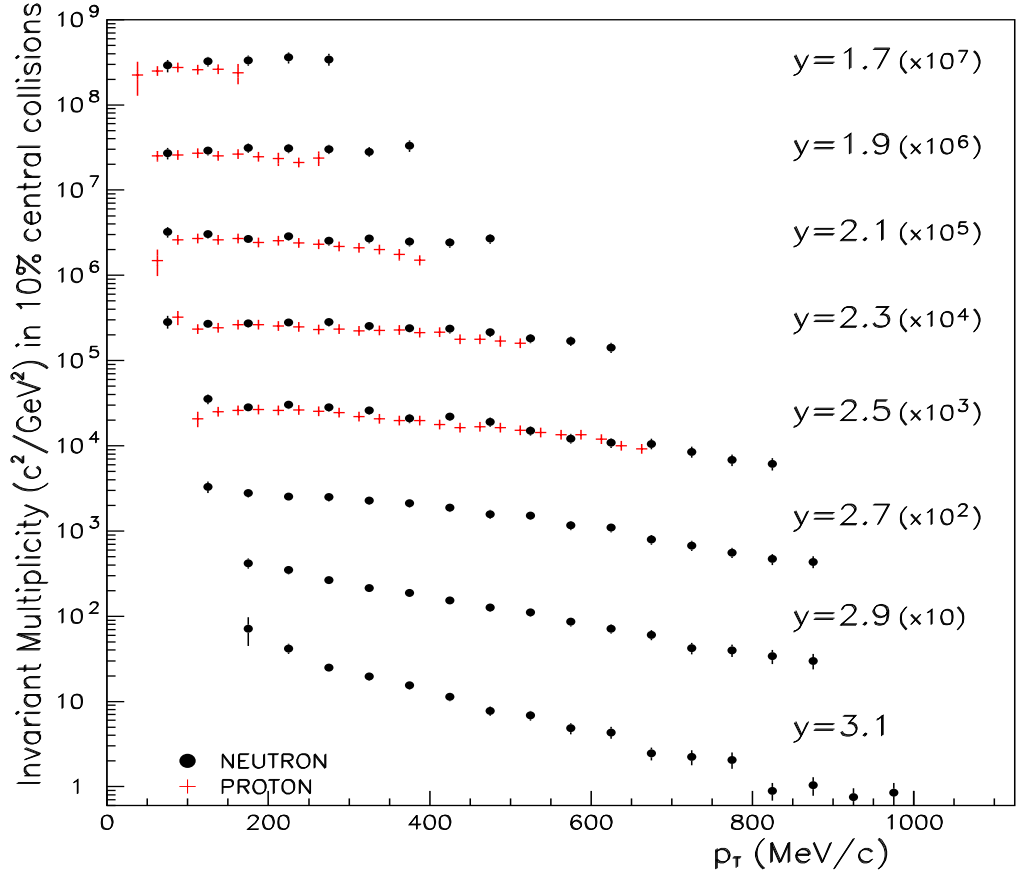}
\caption{Neutron invariant multiplicities in 10\% most central Au+Pb collisions
as a function of transverse momentum 
for rapidities from near center of mass ($y$=1.6) to beam ($y$=3.2).  Each 
rapidity range is multiplied by a successive factor of ten for presentational 
purposes.  Proton multiplicities measured by E864 in central collisions 
are shown for comparison for 
rapidities 1.7 to 2.5.}
\label{fig:inv_mult_np_2}
\end{figure}

Measurements of neutron multiplicity in 10\% most central Au~+~Pb collisions
are presented in Figure~\ref{fig:inv_mult_np_2} and in Table~\ref{tab:cnt}. 
Also shown in Figure~\ref{fig:inv_mult_np_2} are E864 
measurements of proton invariant multiplicity \cite{Nigel} in central 
collisions for the rapidity range where they overlap the neutron measurements.  
Measurements in each rapidity bin are multiplied by a different
factor of ten for presentational purposes.  Agreement with 
the protons is quite close where
comparisons are possible.  This is consistent with the assumption that
the spectra of the two species should not differ considerably other than by 
an overall scale factor, justifying for example the calculation of the  
light nuclei coalescence parameters, $B_{A}$, in terms of the ratio 
of coalesced nuclei to protons only
rather than protons and neutrons. 

\begin{figure}
\centering\epsfxsize=5in \epsfysize=6in \leavevmode \epsfbox{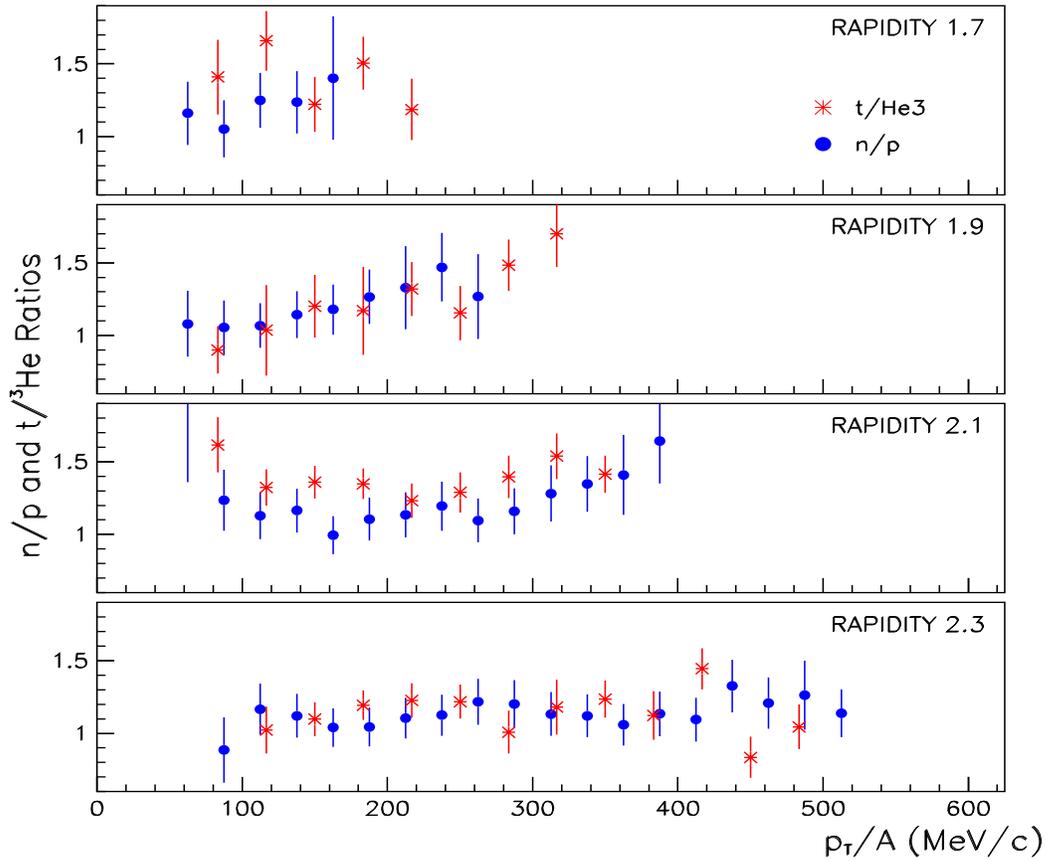}
\caption{Ratios of neutron to proton and triton to $^{3}He$ invariant 
multiplicities in 10\% central collisions plotted versus transverse momentum 
per nucleon $p_{T}/A$ for the rapidity range where all four species are 
measured by E864.  Corrections for feed down from hyperon states are not 
included in the $n/p$ ratios shown here; see the text for discussion of
this point.}
\label{fig:ratios_2}
\end{figure}

The agreement between the two species is made more quantitative 
by Figure~\ref{fig:ratios_2}   
in which we show the neutron to proton ratio along with the triton to 
$^{3}He$ ratio \cite{Nigel} in the rapidity region where all four 
species are 
measured by E864.  If we assume no kinematic dependence of the $n/p$ 
ratio and take a statistically weighted average of each point shown in 
Figure~\ref{fig:ratios_2}, we find an average $n/p$ ratio of $1.14 \pm .08$.  

To determine the ratio which is present at hadronic freeze-out
of the system, we need to subtract from the nucleon multiplicities
the results of feed-down from hyperon decays which occur long after
freeze-out.
To make this subtraction, we assume a $\Lambda$ distribution 
which is parameterized according to measurements
by E891 \cite{E891lambda} and use a distribution for the $\Sigma$ hyperons 
given by the cascade code RQMD version 2.3.  We follow the hyperon decay 
products through a GEANT simulation of the E864 apparatus to determine the 
number of neutrons and protons in each momentum bin which are produced in 
these decays.  We find that the contribution to nucleon invariant 
multiplicities from hyperon feed down is
on the order of 15\% .  After correcting for this feed down, we find 
an average $n/p$ ratio at freeze out of $1.19 \pm .08$.

At rapidities far from the beam rapidity of 3.2 such as are 
shown in Figure~\ref{fig:ratios_2}, light nuclei are very 
unlikely to be beam fragments \cite{dq1,dq2} and must therefore be 
formed by a 
coalescence mechanism.  We expect then that the triton to $^{3}He$ ratio 
should match the neutron to proton ratio which is present at the time when this
coalescence occurs.  Computing a statistically weighted average, we 
obtain a $t/^{3}He$ ratio of $1.23 \pm .04$. which is consistent with our
value for the freeze out $n/p$ ratio.
 
\begin{figure}
\centering\epsfxsize=5in \epsfysize=6in \leavevmode \epsfbox{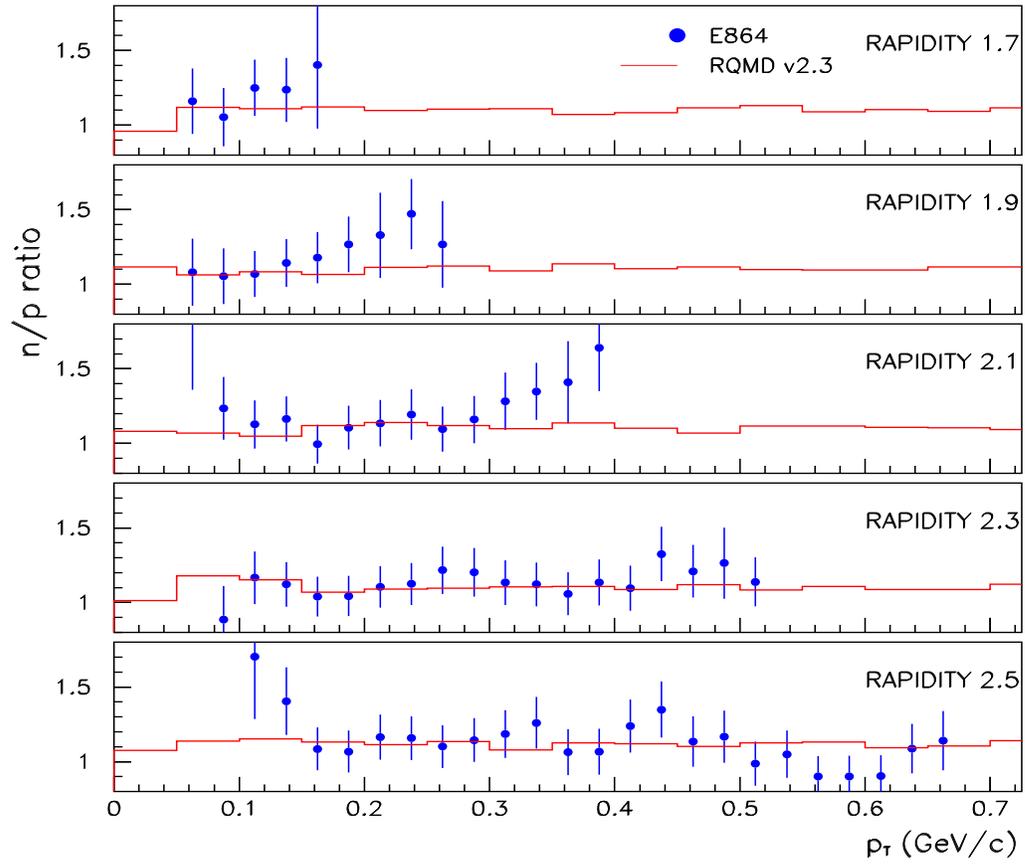}
\caption{Neutron to proton ratio in 10\% central collisions and predictions
of the ratio by RQMD version 2.3 without mean field potentials. The E864 ratios
shown here have not been corrected for feed down from hyperon states; see the 
text for a discussion of this point.}
\label{fig:rqmd_ratio_2}
\end{figure}

The incident nuclei have a total $n/p$ ratio of 1.52, so this observed 
final ratio of 1.19 signifies considerable equilibration of the two 
species from their initial abundances.  This is not surprising in light of the 
amount of strong resonance production which is believed to occur in the 
collision system and which should facilitate the evolution 
of the system toward chemical equilibrium.  Evidence for a large amount 
of $\Delta(1232)$ resonance
production is present in the measured pion transverse momentum spectra and 
$\pi^{-}$ to $\pi^{+}$ ratio in Au~+~Au type collisions at the AGS from 
experiments 866 \cite{E866} and 877\cite{E877}.  
RQMDv2.3 in fact predicts 
that for some duration of the evolution of an AGS Au~+~Au collision, the 
majority of baryons exist as strong resonances \cite{hofmann} and the 
resulting RQMD prediction for n/p ratios match reasonably well with 
our measurements (see Figure~\ref{fig:rqmd_ratio_2}).  
Indeed, in a simplified isobar model (following \cite{pelte}) in which 
half of all the
incident nucleons are excited to resonances by the reaction
$N + N \rightarrow N + \Delta$ with isospin conserved, the neutron
to proton ratio reaches a value of less than 1.1 without any further
interactions.

In a model which imposes chemical and thermal equilibrium on the system, 
there is the approximate constraint that 
$R_{1} \equiv (N_{n}/N_{p})^{2} = (N_{\pi^{-}} / N_{\pi^{+}}) $ (this
is only strictly true if we assume also a Boltzmann distribution for
each species).  
If we also impose the approximate conditions that the total number
of nucleons is conserved (ignoring strange baryons) and that the 
total charge of the nucleons
plus pions at freeze-out is equal to the initial charge of the system, 
we can obtain an approximate value
for the ratio $R_{2} \equiv (N_{\pi^{+}} + N_{\pi^{-}}) / (N_{n} + N_{p})$.
With a freeze out neutron to proton ratio of 1.19, we obtain values
of approximately 1.4 and 3 for $R_{1}$ and $R_{2}$, respectively.  
Including feed-down from resonances in this simple equilibrium model does
little to change these numbers.  

These ratios are strongly
dependent on the input $n/p$ ratio, and the results of these simple calculations
can be made to agree reasonably well with measurements by E866 
\cite{E866} and E877 \cite{E877} if we instead 
assume an $n/p$ ratio of 1.11 (the lower end of the range included in 
1.19$\pm$.08).  This 
yields values of approximately 1.2 and 1.3 for $R_{1}$ and $R_{2}$, 
respectively; thus this set of measurements can be accommodated 
within this simple model.  Note also that the inclusion of a light 
quark saturation factor of 
larger than 1 as proposed in \cite{rafelski} can change the
predictions from such a simple picture. 

\begin{figure}
\centering\epsfxsize=5in \epsfysize=6in \leavevmode \epsfbox{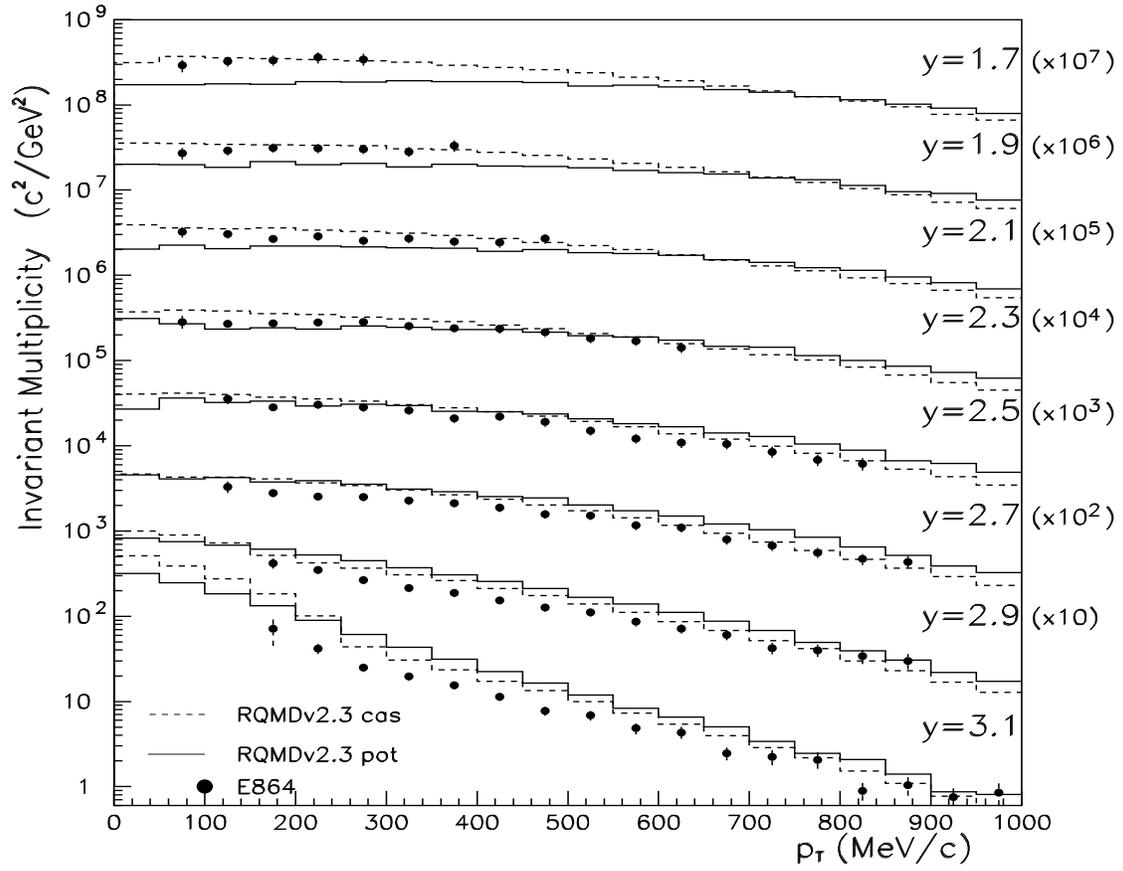}
\caption{Neutron multiplicities in 10\% central collisions and predictions of
RQMD version 2.3 with (solid histogram) and without (dashed histogram) mean 
field potentials.}
\label{fig:rqmd_dndydpt}
\end{figure}

The predictions of neutron invariant multiplicity from 
RQMD version 2.3 \cite{RQMD} with and without
mean field potentials are shown in Figure~\ref{fig:rqmd_dndydpt}.  
As demonstrated, with potentials turned on the multiplicities near center of 
mass rapidity are under predicted by a factor of
approximately two.  Agreement in this rapidity range is much better with mean 
fields switched off.  Near beam rapidity,
RQMD over predicts the neutron yields; this is due at least in part to the fact 
that light nuclei are not included in RQMD but are likely
present in large numbers as beam fragments near beam rapidity.

\begin{figure}
\centering\epsfxsize=5in \epsfysize=6in \leavevmode \epsfbox{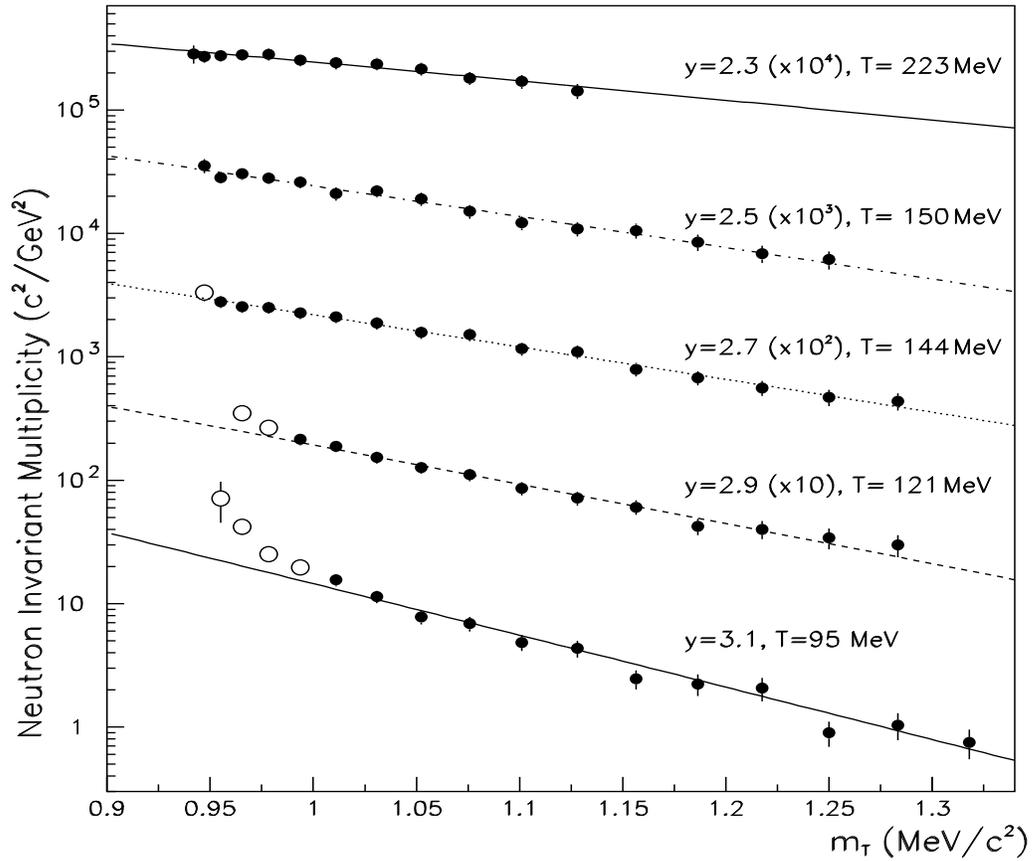}
\caption{Neutron invariant multiplicities plotted versus transverse mass
and fit to a Boltzmann function in each rapidity slice to extract the inverse
slope parameter, $T$, as a function of rapidity.  
Points at low $p_{T}$ near beam
rapidity (represented by hollow circles) were excluded from the fits to
minimize the contribution of spectators.}
\label{fig:boltz_2}
\end{figure}

In the rapidity bins in which we have sufficient coverage in transverse 
momentum, the neutron data fit well to Boltzmann distributions in transverse 
mass,
\begin{equation}
\frac{1}{2\pi p_{T}} \frac{d^{2}N}{dydp_{T}} \propto m_{T}e^{-\frac{m_{T}}{T}}
\end{equation} 
(with $m_{T} = \sqrt{p_{T}^{2}+m^{2}}$) as shown in Figure~\ref{fig:boltz_2}.  
The extracted inverse slope parameters, $T$, are shown in 
Table~\ref{tab:alltemps}.  For the fits shown in Figure~\ref{fig:boltz_2},
we have excluded 
the points at lowest transverse 
momentum near beam rapidity (shown in Figure~\ref{fig:boltz_2} 
as hollow circles) to minimize the effect of spectator neutrons 
on these slope parameters.  Alternatively, we can use a fit to a sum of two 
Boltzmann distributions in these bins to account for these spectator neutrons, 
and the resulting slope parameters are the same as shown in 
Table~\ref{tab:alltemps} within the quoted uncertainties.  
For the bins $y$=2.3 and 
$y$=2.5 where we also have measurements of proton inverse slope parameters, 
the slopes agree quite closely (see Table~\ref{tab:alltemps}). 

\begin{figure}
\centering\epsfxsize=5in \epsfysize=6in \leavevmode \epsfbox{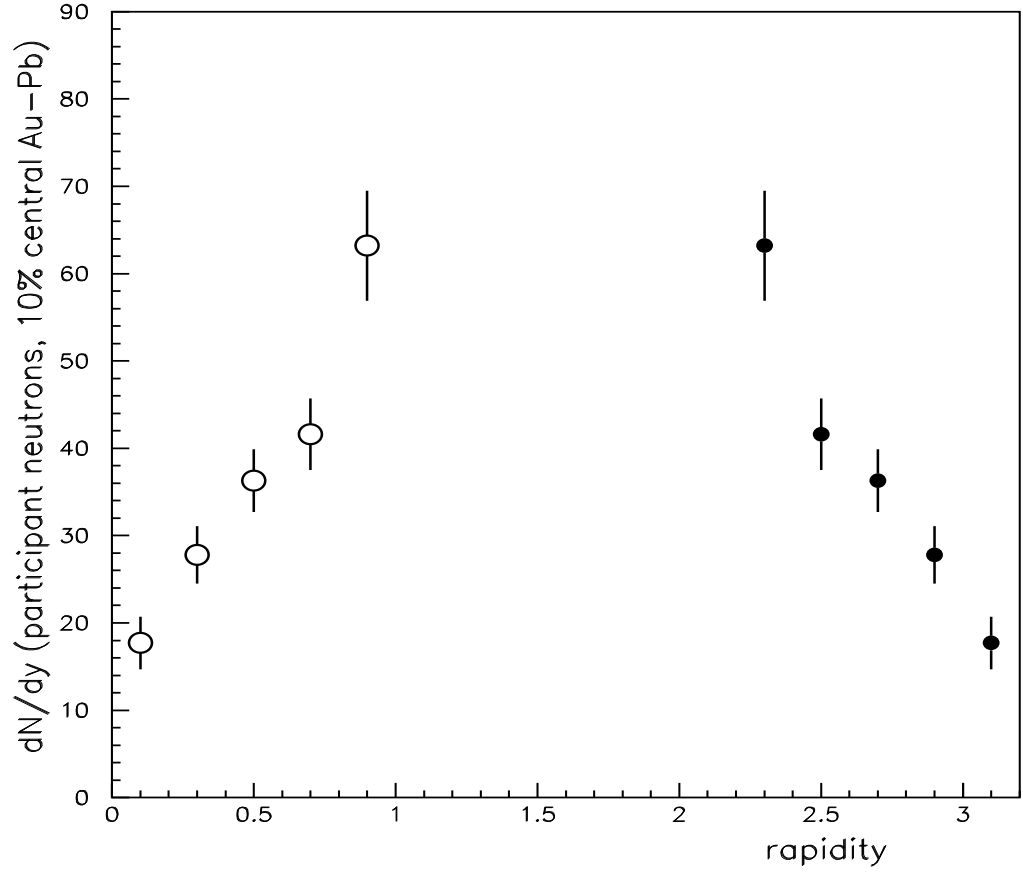}
\caption{dN/dy measured for 'participant' neutrons in central Au+Pb
collisions.  For those $m_{T}$ ranges not covered by direct measurements
(or having a large contribution from spectator neutrons), a Boltzmann fit
is used for the integration in $m_{T}$.  Hollow points denote a reflection
about center-of-mass rapidity.}
\label{fig:dndy}
\end{figure}

In Figure~\ref{fig:dndy} we display the yields dN/dy for participant neutrons.
These were determined by directly integrating our measurements where available
and extrapolating with the Boltzmann fits shown in Figure~\ref{fig:boltz_2}
where necessary.  The points with significant contributions from spectator
neutrons which are displayed in Figure~\ref{fig:boltz_2} as
hollow points were not integrated directly (i.e. the
Boltzmann fit was used to integrate these points).

\begin{figure}
\centering\epsfxsize=5in \epsfysize=6in \leavevmode \epsfbox{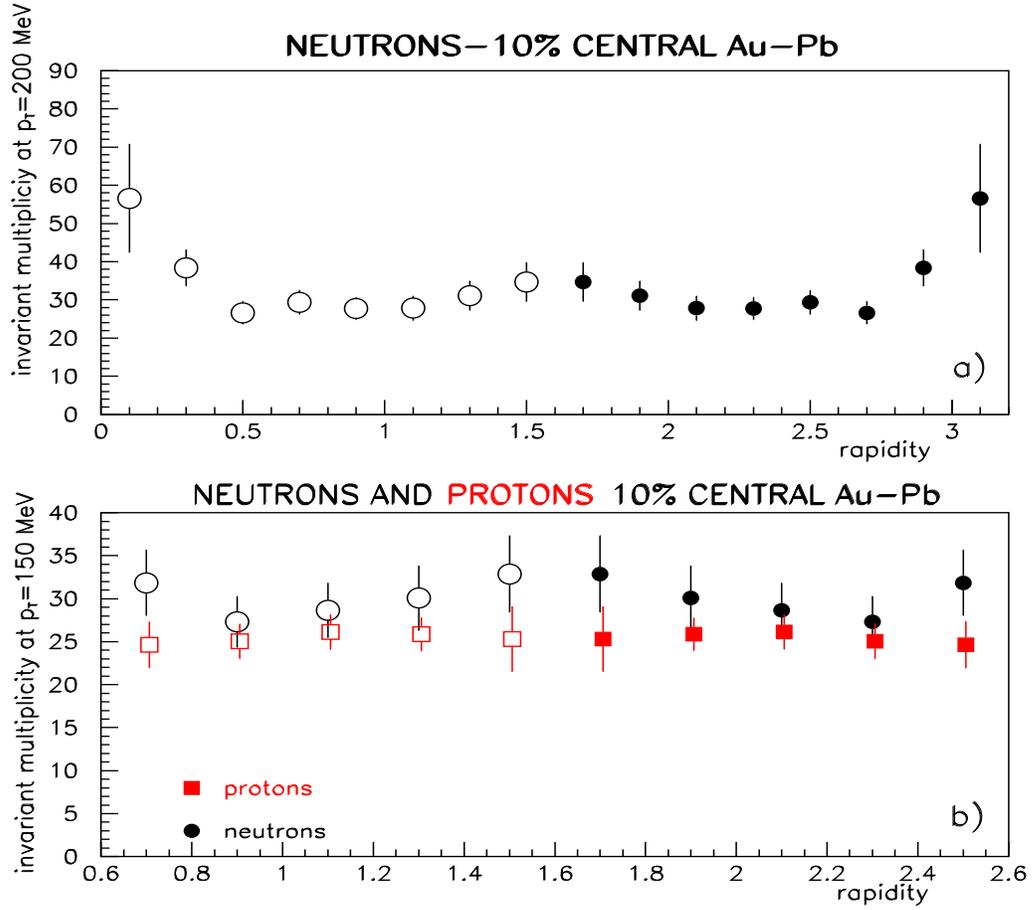}
\caption{Panel a) shows neutron invariant multiplicity in a transverse momentum
range from 150 to 250 MeV/c
as a function of rapidity.  Panel b) shows a slightly different $p_{T}$ range 
(100 to 200 MeV/c) and covers a smaller range in 
rapidity for comparison with proton 
invariant multiplicity.  Hollow symbols denote reflections of measurements
about center-of-mass rapidity.}
\label{fig:pt150and200_2}
\end{figure}

Due to our limited coverage in $p_{T}$ near center of mass rapidity we cannot 
accurately integrate the $m_{T}$ spectra to measure $dN/dy$ for neutrons in this
region.  To 
examine the behaviour of the spectrum as a function of rapidity we plot the 
invariant multiplicity near $p_{T}$=0 versus $y$ in Figure 
~\ref{fig:pt150and200_2} a).  This $p_{T}$ range (150 to 250 MeV/c) was chosen 
because it was common among all rapidity bins.  A similar plot showing 
comparison with the protons in a similar $p_{T}$ bite (100 to 200 MeV/c) is 
shown in Figure~\ref{fig:pt150and200_2} b).  
There is some evidence here that the 
neutrons exhibit a slight peak near midrapidity while the protons are flat, but
in light of the size of the systematic errors on these points, 
this evidence is slight.  
One can ask if such a difference in shape would be consistent
with the additional coulomb repulsion felt by the protons.  Under the 
assumption that the coulomb force only has an effect after the 
nucleons reach freeze-out from the strong force, one can with a very 
simplified model estimate 
the effect of the 
coulomb force on a proton following reference \cite{baym}.  Assuming a freeze 
out radius $r$ and that
all net charge of the source is contained within $r$ (in a simple 
spherically symmetric model, this should provide a generous upper limit 
for the 
coulomb effect), a proton with center of mass momentum $p_{p}$ will be 
accelerated to a momentum of 
$\sqrt{p_{p}^{2}+ 2 Z_{N} e^{2}/r }$.  Taking $Z_{N}$ = 150 and $r$ = 5 fm,
we find that a proton at center of mass rapidity with $p_{T}$=150 will
receive an extra $p_{T}$ kick of 20 MeV/c from the coulomb interaction. 
With a more realistic assumption including some form of radial flow, 
however, the amount of charge that is contained within
a sphere with radius equal to the freeze out radius of such a 
proton should be at most only a few percent of $Z_{N}$ and so we do 
not expect any observable 
effect from the coulomb interaction.

\begin{figure}
\centering\epsfxsize=5in \epsfysize=6in \leavevmode \epsfbox{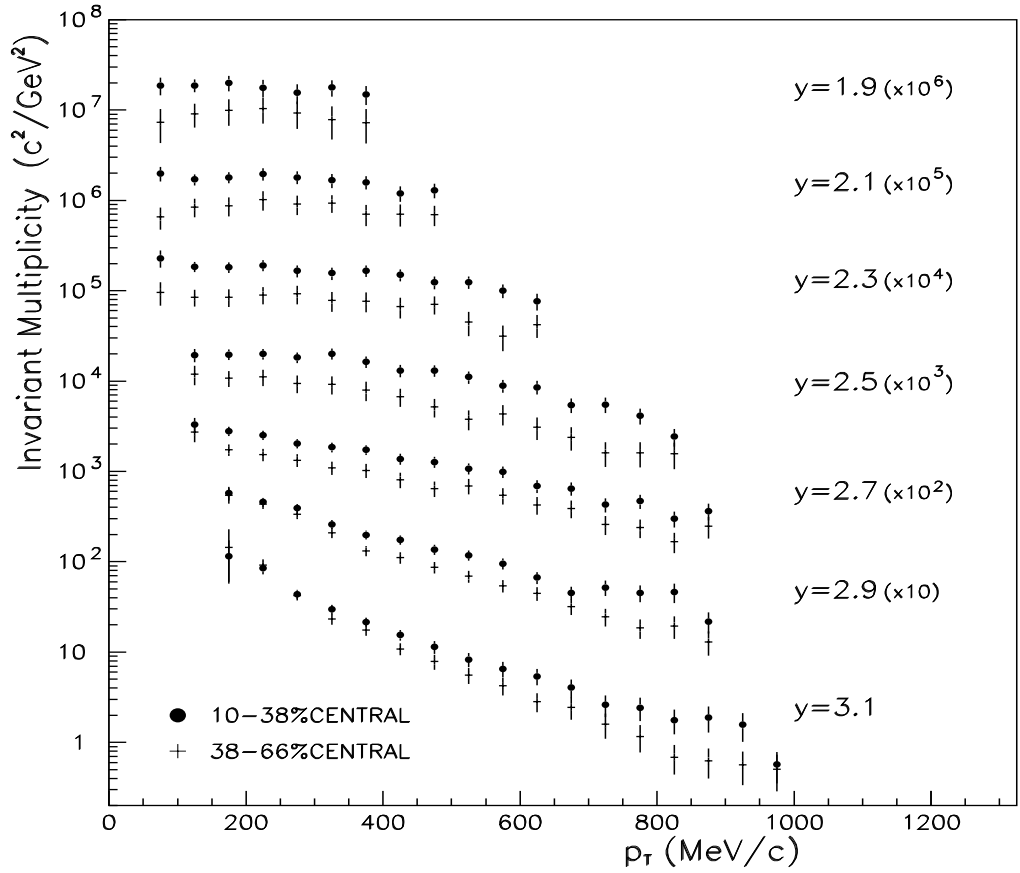}
\caption{Neutron invariant multiplicities in less central events plotted versus
transverse momentum.  Shown as the solid
circles are multiplicities in 10-38\% most 
central events and shown as the crosses
are multiplicities in 38-66\% most central 
events (the most peripheral collisions
for which we make neutron measurements).}
\label{fig:other_cent}
\end{figure}

Shown in Figure~\ref{fig:other_cent} are the neutron multiplicities for 
10-38\% and 38-66\% most central events.  These measurements include 
larger uncertainties than are present in the 10\% most central data, 
particularly near center of mass rapidity where to a first approximation 
the neutron signal scales as the number of participants while 
background sources tend to remain constant or grow as the number of spectators.
Corrections due to beam interactions which do not occur in the target are taken
into account for these centralities using data from empty target runs, and 
this is not a significant source of systematic error.

We do see the qualitative behaviour which we expect in these centralities;  the
multiplicities near center of mass rapidity scale crudely with the number
of participant nucleons and we see larger contribution from spectator neutrons
at high rapidity and low $p_{T}$ as we go to less central events.
We also note that the inverse slope parameters (see Table~\ref{tab:alltemps}) 
become smaller as centrality decreases and as in the central data agreement 
between the proton and neutron inverse slope parameters is quite close where
comparisons are possible.

\section{Summary}

We have presented results from Experiment 864 for neutron invariant 
multiplicities produced in 11.5 A GeV/c Au~+~Pb collisions.  
These are the first 
neutron measurements for a system of comparable size at AGS energies or above.  

We observe little kinematic dependence of the neutron to proton ratio, 
consistent with the idea that the neutron spectrum should to a good 
approximation differ from the proton spectrum only by an overall scale factor.  
An average neutron to proton freeze-out 
ratio of $1.19 \pm .08 $ is observed within .8 units of midrapidity.  This 
value is consistent with E864 measurements of the ratio of coalesced tritons 
to $^{3}He$ nuclei and represents a significant equilibration from the 
initial state. 

\section{Acknowledgements}

We gratefully acknowledge the efforts of the AGS staff in providing the beam.  
This work was supported in part by grants from the Department of Energy 
(DOE) High Energy Physics Division, the DOE Nuclear Physics Division, and 
the National Science Foundation.


\newpage

%
%

%
%
\newpage

\begin{table}
\caption{Summary of backgrounds present under the neutron peak in 
central Au+Pb collisions, along with their approximate level
in two different rapidity ranges (variations with $p_{T}$ within these different
rapidity ranges can be significant) and the method by which this is determined.}
\label{tab:bkg}
\begin{tabular}{|c|c|c|c|}
                       		& Level at $y\approx2.0$
&Level at $y\approx2.8$	& Determined by                        \\ \tableline
 Background type		& 			
& 			&			               \\ \tableline
 Low mass              		&  5-10 \%		
&   1-3 \%		&  Fit to data: mass peak + background \\ \tableline
 $K_{L}$s,$\gamma$s,protons	&  7-11 \%		
&   2-13 \%		&  GEANT+RQMD simulations              \\ \tableline
 Scattered neutrons    		&  8-20 \%	 	
&   4-9 \%		&  GEANT+RQMD simulations, proton data \\ \tableline
 All other particles   		& $<$1 \%		
&  $<$1 \%		&  GEANT+RQMD simulations              \\ 
\end{tabular}
\end{table}  

\begin{table}
\caption{Efficiency for individual analysis cuts in two different momentum
regions, both for neutrons showers amidst the occupancy of a central 
heavy ion event (labelled: CENTRAL) and for neutrons showers 
in an otherwise empty 
calorimeter (labelled: EMPTY).  Values for transverse 
momentum are given in MeV/c.}
\label{tab:indiecuts}
\begin{tabular}{|c|c|c|c|c|}
 		       & \multicolumn{2}{c|}{$1.8<y<2.2,150<p_{T}<250 $}
& \multicolumn{2}{c|}{$2.6<y<3.0, 400<p_{T}<550 $}\\ \tableline \tableline
                       & EMPTY                  & CENTRAL		
& EMPTY			& CENTRAL                 \\ \tableline
 Analysis Cut	       &                        & 			
&			&                         \\ \tableline
 $t_{max}$             &    98 \%               &  62 \%		
&   93 \%		& 76 \%                   \\ \tableline
 $E_{PK}$              &   100 \%		&  74 \%		
&  100 \%		& 75 \%                   \\ \tableline
 $R_{5x5/3x3}$         &   100 \%		&  78 \% 		
&  100 \%		& 94 \%                   \\ \tableline
 Charged particle veto &    93 \%		&  93 \%		
&   93 \%		& 93 \%                   \\ \tableline
 Shower Energy Profile &    89 \%		&  66 \%	   	
&   85 \%		& 68 \%                   \\ \tableline
 $.55<mass<1.55 GeV/c$ &    92 \%		&  71 \%   		
&   80 \%		& 68 \%                   \\
\end{tabular}
\end{table}  

\begin{table}
\caption{Neutron invariant multiplicities for 10\% most central Au~+~Pb 
events and their total uncertainties (statistical and systematic errors 
are added in 
quadrature.)}
\label{tab:cnt}
\begin{tabular}{ccccccccc}
 $y$     &1.7      & 1.9     & 2.1    & 2.3    & 2.5    & 2.7    & 2.9    & 3.1
    \\ \tableline \tableline
 $p_{T}(MeV)$ &         &         &        &        &        &        &        
&        \\ \tableline
 75 & 29.0$\pm$  4.8 & 27.2$\pm$  4.2 & 32.3$\pm$  4.6 & 28.5$\pm$  4.8 &      
          &                &                &                \\ \tableline
125 & 32.3$\pm$  4.2 & 29.0$\pm$  3.6 & 30.5$\pm$  3.3 & 27.1$\pm$  3.0 & 35.4$
\pm$  4.7 & 33.1$\pm$  5.1 &                &                \\ \tableline
175 & 33.4$\pm$  4.7 & 31.1$\pm$  3.9 & 26.8$\pm$  3.0 & 27.5$\pm$  2.9 & 28.3$
\pm$  3.0 & 27.9$\pm$  3.1 & 41.9$\pm$  5.9 & 71.3$\pm$ 26.3 \\ \tableline
225 & 36.0$\pm$  5.4 & 31.0$\pm$  3.8 & 28.8$\pm$  3.3 & 28.1$\pm$  2.9 & 30.3$
\pm$  3.2 & 25.4$\pm$  2.8 & 34.8$\pm$  3.8 & 41.8$\pm$  5.5 \\ \tableline
275 & 34.2$\pm$  5.6 & 30.2$\pm$  3.7 & 25.4$\pm$  2.9 & 28.2$\pm$  3.1 & 28.1$
\pm$  3.0 & 25.0$\pm$  2.8 & 26.6$\pm$  2.9 & 25.1$\pm$  2.7 \\ \tableline
325 &                & 28.2$\pm$  3.7 & 27.0$\pm$  3.1 & 25.3$\pm$  2.8 & 26.1$
\pm$  2.9 & 22.7$\pm$  2.5 & 21.4$\pm$  2.3 & 19.6$\pm$  2.2 \\ \tableline
375 &                & 33.2$\pm$  5.1 & 24.7$\pm$  2.9 & 24.1$\pm$  2.7 & 21.0$
\pm$  2.6 & 21.1$\pm$  2.4 & 18.9$\pm$  2.0 & 15.5$\pm$  1.8 \\ \tableline
425 &                &                & 24.2$\pm$  3.1 & 23.6$\pm$  2.7 & 22.0$
\pm$  2.5 & 18.7$\pm$  2.1 & 15.3$\pm$  1.7 & 11.4$\pm$  1.3 \\ \tableline
475 &                &                & 27.0$\pm$  3.7 & 21.5$\pm$  2.6 & 19.0$
\pm$  2.3 & 15.8$\pm$  1.8 & 12.7$\pm$  1.4 &  7.8$\pm$  1.0 \\ \tableline
525 &                &                &                & 18.1$\pm$  2.1 & 15.1$
\pm$  1.9 & 15.2$\pm$  1.8 & 11.1$\pm$  1.3 &  6.9$\pm$  0.9 \\ \tableline
575 &                &                &                & 17.0$\pm$  2.1 & 12.1$
\pm$  1.6 & 11.6$\pm$  1.4 &  8.6$\pm$  1.1 &  4.8$\pm$  0.7 \\ \tableline
625 &                &                &                & 14.2$\pm$  2.0 & 10.9$
\pm$  1.4 & 11.0$\pm$  1.3 &  7.1$\pm$  0.9 &  4.3$\pm$  0.7 \\ \tableline
675 &                &                &                &                & 10.5$
\pm$  1.5 &  7.9$\pm$  1.0 &  6.0$\pm$  0.8 &  2.5$\pm$  0.4 \\ \tableline
725 &                &                &                &                &  8.5$
\pm$  1.3 &  6.8$\pm$  0.9 &  4.2$\pm$  0.6 &  2.2$\pm$  0.4 \\ \tableline
775 &                &                &                &                &  6.8$
\pm$  1.1 &  5.6$\pm$  0.8 &  4.0$\pm$  0.7 &  2.1$\pm$  0.4 \\ \tableline
825 &                &                &                &                &  6.1$
\pm$  1.0 &  4.7$\pm$  0.7 &  3.4$\pm$  0.6 &  0.9$\pm$  0.2 \\ \tableline
875 &                &                &                &                &      
          &  4.3$\pm$  0.7 &  3.0$\pm$  0.6 &  1.0$\pm$  0.3 \\ \tableline
925 &                &                &                &                &      
          &                &                &  0.8$\pm$  0.2 \\ \tableline
975 &                &                &                &                &      
          &                &                &  0.9$\pm$  0.3 \\ 
\end{tabular}
\end{table}  

\begin{table}
\caption{Inverse slope parameters extracted from Boltzmann fits to neutron 
($T_{n}$) and 
proton ($T_{p}$) spectra. Shown in parentheses after 
each value for $T_{n}$ is the 
total $\chi^{2}$ / number of degrees of freedom of the Boltzmann fit from which 
these values were extracted; these reduced $\chi^{2}$ 
values are generally significantly less than one due to 
the presence of some amount 
of common mode uncertainty in each rapidity bin which will not affect the 
extracted $T_{n}$.}
\label{tab:alltemps}
\begin{tabular}{|c|c|c|c|c|c|}
 Rapidity                   & 2.3                &  2.5               
&  2.7               & 2.9                 & 3.1       \\ \tableline \tableline
 $T_{n}$(MeV) 10\% central  &223$\pm$27 (2.3/10) &150$\pm$11 (4.1/13) 
&144$\pm$10 (2.6/13) &121$\pm$11 (3.7/10) &95$\pm$10 (14/11)     \\ \tableline
 $T_{p} $(MeV) 10\% central &213$\pm$16          &167$\pm$10          
&                    &                    &                  \\ \tableline
 $T_{n}$(MeV) 10-38\%       &189$\pm$24 (2.8/10) &138$\pm$11 (7.3/13) 
&129$\pm$9 (6.9/13)  &112$\pm$10 (10.6/10)&95$\pm$10 (8.7/11)     \\ \tableline
 $T_{p} $(MeV) 10-38\%      &175$\pm$12          &                    
&                    &                    &                  \\ \tableline
 $T_{n}$(MeV) 38-66\%       &172$\pm$30 (3.2/10) &119$\pm$12 (2.4/13) 
&126$\pm$11 (6.6/13) & 95$\pm$11 (7.1/10) &80$\pm$9  (9.5/11)    \\ \tableline
 $T_{p} $(MeV) 38-66\%      &143$\pm$8           &                    
&                    &                    &                  \\ 
\end{tabular}
\end{table}


\begin{references}
\bibitem[\ast]{}        Present address: Vanderbilt University, 
Nashville, Tennessee 37235 
\bibitem[\dag]{}        Present Address: Anderson Consulting, Hartford, CT
\bibitem[\ddag]{} 	Present address: Univ. of Denver, Denver CO 80208
\bibitem[\S]{}    	Deceased.
\bibitem[\|]{}     	Present address: McKinsey \& Co., New York, NY 10022
\bibitem[\P]{}          Present address: Department of Radiation Oncology, 
Medical College of Virginia, Richmond VA 23298
\bibitem[\ast\ast]{}    Present address: University of Tennessee, 
Knoxville TN 37996
\bibitem[\dag\dag]{}  Present address: Institut de Physique 
Nucl\'{e}aire, 91406 Orsay Cedex, France
\bibitem[\ddag\ddag]{} 	Present Address: Institute for Defense 
Analysis, Alexandria VA 22311
\bibitem[\S\S]{} 	Present Address: MIT Lincoln Laboratory, 
Lexington MA 02420-9185
\bibitem{RQMD} H. Sorge et al., Nucl. Phys. A525(1991) 95.
\bibitem{E866} L. Ahle et al., Nucl. Phys. A610(1996) 139.; Z. Chen et al., 
 in Proceedings of HIPAGS '96, Report WSU-NP-96-16.
\bibitem{E877} R. Lacasse et al., Nucl. Phys. A610(1996) 153.; T. W. Piazza et
 al. in, Proceedings of HIPAGS '96, Report WSU-NP-96-16.
\bibitem{Nigel} T. A. Armstrong et. al., to be submitted to Phys. Rev. C.;
N. K. George, PhD Thesis, Yale University 1999.;
T. A. Armstrong et. al., submitted to Phys. Rev. Lett., nucl-ex/9907002.
\bibitem{jamie_col} J.L. Nagle et. al., Phys. Rev. C 53(1996) 367.
\bibitem{sato} H. Sato and K. Yazaki, Phys. Lett. 98B{(1981)} 153.
\bibitem{sis} Y. Leifels et al., Phys. Rev. Lett. 71{(1993)} 963.
\bibitem{bignim}T. A. Armstrong et al., accepted for publication 
 in Nucl. Inst. Meth.
\bibitem{claudecal}T. A. Armstrong et al., Nucl. Inst. Meth. A406(1998)227.
\bibitem{beam} P. Haridas et al., Nucl. Inst. Meth. A385(1997) 413.
\bibitem{evanthes} L. Evan Finch, PhD Thesis, Yale University 1999.
\bibitem{xzbthes} Z. Xu, PhD Thesis, Yale University 1999.
\bibitem{stankus} The WA98 Collaboration, internal document "Photon Efficiency
 Calculation Using Overlap Method".
\bibitem{E891lambda} S. Ahmad et al., Phys. Lett. B 382(1996) 35.
\bibitem{dq1} L. C. Alexa et al., Phys. Rev. Lett. 82{(1999)} 1374;
\bibitem{dq2} D. Abbott et al., Phys. Rev. Lett. 82{(1999)} 1379.
\bibitem{hofmann} M. Hofmann et al., Phys. Rev. C 51(1995) 2095.
\bibitem{pelte} D. Pelte.  Talk presented at Park City Utah, 
 January 9-16, 1999. nucl-ex/9902006.
\bibitem{rafelski} J. Rafelski and J. Letessier.  
 Talk presented at Park City Utah, January 9-16, 1999. hep-ph/9902365.
\bibitem{baym} G. A. Baym and P. Braun-Munzinger, Nucl. Phys. A610(1996) 286.

\end{references}
\end{document}